\documentclass[11pt]{article}%
\usepackage{hyperref}
\usepackage{latexsym}
\usepackage{amssymb,amsfonts,amsmath}
\usepackage{graphicx}
\usepackage{indentfirst}
\usepackage{amsmath}
\usepackage{amsfonts}
\usepackage{amssymb}%
\ifx\pdfoutput\undefined
\else
\fi
\hypersetup{colorlinks=false,bookmarksopen,bookmarksnumbered,citecolor=blue,
pdfstartview=FitH}

\parskip=\medskipamount

\newcommand{\be}{\begin{equation}}
\newcommand{\ee}{\end{equation}}
\newcommand{\bea}{\begin{eqnarray}}
\newcommand{\eea}{\end{eqnarray}}

\newcommand{\ba}{\begin{array}}
\newcommand{\ea}{\end{array}}

\def\double #1{#1{\hbox{\kern-2pt $#1$}}}

\newcommand{\bsubeq}{\begin{subequations}}
\newcommand{\esubeq}{\end{subequations}}

\begin{document}

\begin{titlepage}
\begin{flushright}
ARC-18-01
\par\end{flushright}
\vskip 1.5cm
\begin{center}
\textbf{\huge \bf String Sigma Models}\\\vspace{.2cm}
\textbf{\huge \bf on Curved Supermanifolds}

\vskip 1.5cm
{\large
Roberto Catenacci$^{~a,b,c,}$\footnote{roberto.catenacci@uniupo.it}
and
Pietro Antonio Grassi$^{~a,b,d,}$\footnote{pietro.grassi@uniupo.it}
\medskip
}
\vskip 0.5cm
{
\small\it
\centerline{${(a)}$ Dipartimento di Scienze e Innovazione Tecnologica, Universit\`a del Piemonte Orientale} }
\centerline{\it Viale T. Michel, 11, 15121 Alessandria, Italy}
\medskip
\centerline{${(b)}$ {\it
Gruppo Nazionale di Fisica Matematica, INdAM, P.le Aldo Moro 5, 00185 Roma, Italy} }
\medskip
\centerline{${(c)}$ {\it Arnold-Regge Center, via P. Giuria 1, 10125 Torino, Italy}}
\medskip
\centerline{${(d)}$ {\it
INFN, Sezione di Torino, via P. Giuria 1, 10125 Torino, Italy} }
\vskip  .5cm
\medskip
\par\end{center}
\vfill{}
\begin{abstract}
We use the techniques of integral forms to analyse the easiest example of two dimensional sigma models 
on a supermanifold. We write the action as an integral of a top integral form over a D=2 supermanifold and we show 
how to interpolate between different superspace actions. Then, we consider curved supermanifolds
and we show that the definitions used for flat supermanifold can also be used for curved supermanifolds. 
We prove it by first considering the case of a curved rigid supermanifold and then the case of a generic curved supermanifold 
described by a single superfield $E$.
\end{abstract}
\vfill{}
\vspace{1cm}
\centerline{\em Contribution to the Special Issue of Universe ``Super Geometry for Super Strings"}

\end{titlepage}
\newpage\setcounter{footnote}{0}
\tableofcontents
\newpage

\section{Introduction}

During the last years, some new structures in the geometry of supermanifolds
have been uncovered. The conventional\footnote{This is usually called the
bundle of superforms, generated by direct sum of exterior products of
differential forms on the supermanifold.} exterior bundle of a supermanifold
$\Lambda^{\ast}[\mathcal{M}]$ is not the complete bundle needed to construct a
geometric theory of integration, the Hodge dual operation and to study the
cohomology. One has to take into account also the complexes of pseudo and
integral forms.

We call the complete bundle the \textit{integral superspace}. In the present
notes, we consider some new developments and we study also the case of curved
integral superspace.

One of the key ingredient of string theory (in the Ramond-Nevue-Schwarz
formulation) is the worldsheet supersymmetry needed to remove the unphysical
tachyon from the spectrum, to describe fermionic vertex operators and to
construct a supersymmetric spectrum. All of these properties are deeply
related to the worldsheet supersymmetry and they are clearly displayed by
using the superspace approach in $2$ dimensions.

Pertubartive string theory is described by a non-linear sigma model for maps
\begin{eqnarray}
\label{actA}
\phi^{m}(z,\bar{z}),\lambda^{m}(z,\bar{z}),\bar{\lambda}^{m}(z,\bar{z})
\end{eqnarray}
from the worldsheet Riemann surface $\Sigma^{(1)}$ (with one complex
dimension) to a $10$ dimensional target space $\mathcal{M}^{(10)}%
,(m=0,\dots,9)$ with an action given by (for a flat surface $\Sigma
=\mathbb{C}$)
\begin{eqnarray}
\label{actB}
S[\phi,\lambda,\bar{\lambda}]=\int_{\Sigma}d^{2}z\mathcal{L}(\phi,\lambda
,\bar{\lambda})\,.
\end{eqnarray}
Where $\phi,\lambda$ denote respectively the bosonic and the fermionic fields .

To generalize it to any surface, one has to couple the action to $D=2$ gravity
in the usual way, namely by promoting the derivatives to covariant derivatives
and adding the couplings with the $D=2$ curvature. That can be easily done by
considering the action as a $2$-form to be integrated on $\Sigma$ using the
intrinsic definition of differential forms, Hodge duals and the differential
$d$. To avoid using the Hodge dual, one can pass to first order formalism by
introducing some auxiliary fields. Then, we have
\begin{eqnarray}
\label{actC}
S[\phi,\lambda,\bar{\lambda}]=\int_{\Sigma}\mathcal{L}^{(2)}(\phi,\dots
,d\phi,\dots,V^{\pm\pm},\omega)\,.
\end{eqnarray}
where the $2$-form action $\mathcal{L}^{(2)}(\phi,\dots,d\phi,\dots,V^{\pm\pm
},\omega)$ depends upon the fields $(\phi^{m},\lambda^{m},\bar{\lambda}^{m})$,
their differentials $(d\phi^{m},d\lambda^{m},d\bar{\lambda}^{m})$, the $2d$
vielbeins $V^{\pm\pm}$ and the $SO(1,1)$ spin connection $\omega$. To make the
supersymmetry manifest, one can rewrite the action (\ref{actA}) in the
superspace formalism by condensing all fields into a superfield
\begin{eqnarray}
\label{actD}
\Phi^{m}=\phi^{m}+\lambda^{m}\theta^{+}+\bar{\lambda}^{m}\theta^{-}%
+f\theta^{+}\theta^{-}%
\end{eqnarray}
(we introduce the two anticommuting coordinates $\theta^{\pm}$ and their
corresponding derivatives $D_{\pm}$ and the auxiliary field $f$) as follows:
\begin{eqnarray}
\label{actE}
S[\Phi]=\int[d^{2}zd^{2}\theta]\mathcal{L}(\Phi)\,. 
\end{eqnarray}
The integration is extended to the superspace using the Berezin integration
rules. In the same way as above, in order to generalise it to any
\textit{super} Riemann surface $\mathcal{S}\Sigma$ or, more generally, to any
complex $D=1$ supermanifold, we need to rewrite the action (\ref{actE}) as an
integral of an integral form on $\mathcal{S}\Sigma$. As will be explained in
the forthcoming section and as is discussed in the literature
\cite{Castellani:2014goa,Castellani:2015paa,LMP}, the action $\mathcal{L}%
(\Phi)$ becomes a $(2|0)$-superform $\mathcal{L}^{(2|0)}(\Phi,d\Phi,V^{\pm\pm
},\psi^{\pm},\omega)$. It is well wnown that a superform cannot be integrated
on the supermanifold $\mathcal{S}\Sigma$ and it must be converted into an
integral form by multiplying it by a PCO $\mathbb{Y}^{(0|2)}(V^{\pm\pm}%
,\psi^{\pm},\omega)$. The latter is the Poincar\'{e} dual of the immersion of
the bosonic submanifold into the supermanifold $\mathcal{S}\Sigma$. The action
is written as:
\begin{eqnarray}
\label{actF}
S[\Phi]=\int_{\mathcal{S}\Sigma}\mathcal{L}^{(2|0)}(\Phi,d\Phi,V^{\pm\pm}%
,\psi^{\pm},\omega)\wedge\mathbb{Y}^{(0|2)}(V^{\pm\pm},\psi^{\pm},\omega)
\end{eqnarray}
The PCO $\mathbb{Y}^{(0|2)}$ is a $(0|2)$ integral form and is $d$-closed and
not exact. If we shift it by an exact term $\mathbb{Y}+d\Lambda$, the action
is left invariant if $d\mathcal{L}^{(2|0)}(\Phi,d\Phi,V^{\pm\pm},\psi^{\pm
},\omega)=0$. That can be obtained in presence of auxiliary fields and can be
verified using the Bianchi identities for the torsion $T^{\pm\pm}$, the
gravitinos field strengths $\rho^{\pm}$ and the curvature $R$. The choice of
the PCO allows to interpolate between different superspace frameworks with
different manifest supersymmetries.

The action (\ref{actF}) is invariant under superdiffeomorphisms by
construction since it is an intergral of a top integral form. Therefore, it
can be written for any solution of the Bianchi identity for any supermanifold
compatible with them. As will be show in the following, we can write the most
general solution of the Bianchi identities in terms of an unconstrained
superfield $E$.

The paper has the following structure: in sec. 2, we summarize the geometry of
integral forms. In sec. 3, we discuss the PCO's and their properties. In sec.
4, we discuss the action (\ref{actB}) in components and the Bianchi identities
for the field strengths for the superfield $\Phi$. In subsec. 4.1, we derive
the action (\ref{actF}) and we show the relation between the component action
and the superfield action. In sec.~5, we consider the preliminary case of 
rigid curved supermanifold based on the supercoset space ${\rm Osp}(1|2)/SO(1,1)$. 
We show the relation between the volume form and the curvature. In sec.~6, we study the 
general case of 2d supergravity N=1. In particular, it is shown that the PCO in the curved space 
are closed because of the torsion constraints.

\section{Integral forms and integration}

The integral forms are the crucial ingredients to define a geometric
integration theory for supermanifolds inheriting all good properties of
differential forms integration theory in conventional (purely bosonic)
geometry. In the following section we briefly describe the notations and the
most relevant definitions (see \cite{Witten:2012bg}, \cite{voronov-book} and
also \cite{Castellani:2015paa,Castellani:2014goa,Castellani:2015ata}).

We consider a supermanifold with $n$ bosonic and $m$ fermionic dimensions,
denoted here and in the following by $\mathcal{M}^{(n|m)}$ , locally
isomorphic to the superspace $\mathbb{R}^{(n|m)}$. The local coordinates in an
open set are denoted by $(x^{a},\theta^{\alpha})$. When necessary, in sections
4, 5 and 6, we introduce supermanifolds locally isomorphic to the complex
superspace $\mathbb{C}^{(n|m)}$. In this case the coordinates will be denoted
by $(z^{a},\bar{z}^{a},\theta^{\alpha},\bar{\theta}^{\alpha})$; the formalism,
\textit{mutatis mutandis}, is the same.

A $(p|q)$ pseudoform $\omega^{(p|q)}$ has the following structure:
\begin{equation}
\omega^{(p|q)}=\omega(x,\theta)dx^{a_{1}}\dots dx^{a_{r}}d\theta^{\alpha_{1}%
}\dots d\theta^{\alpha_{s}}\delta^{(b_{1})}(d\theta^{\beta_{1}})\dots
\delta^{(b_{q})}(d\theta^{\beta_{q}}) \label{pseudo}%
\end{equation}
where, in a given monomial, the $d\theta^{a}$ appearing in the product are
different from those appearing in the delta's $\delta(d\theta)$ and
$\omega(x,\theta)$ is a set of superfields with index structure $\omega
_{\lbrack a_{1}\dots a_{r}](\alpha_{1}\dots\alpha_{s})[\beta_{1}\dots\beta
_{q}]}(x,\theta)$.

The two integer numbers $p$ and $q$ correspond respectively to the
\textit{form} number and the \textit{picture} number, and they range from
$-\infty$ to $+\infty$ for $p$ and $0\leq q\leq m$. The index $b$ on the delta
$\delta^{(b)}(d\theta^{\alpha})$ denotes the degree of the derivative of the
delta function with respect to its argument. The total picture of
$\omega^{(p|q)}$ corresponds to the total number of delta functions and its
derivatives. We call $\omega^{(p|q)}$ a \textit{superform} if $q=0$ and an
\textit{integral form} if $q=m$; otherwise it is called \textit{pseudoform}.
The total form degree is given by $p=r+s-\sum_{i=1}^{i=q}b_{i}$ since the
derivatives act effectively as negative forms and the delta functions carry
zero form degree. We recall the following properties:
\begin{equation}
d\theta^{\alpha}\delta(d\theta^{\alpha})=0,\text{ }d\delta^{(b)}%
(d\theta^{\alpha})=0\,,~d\theta^{\alpha}\delta^{(b)}(d\theta^{\alpha
})=-b\delta^{(b-1)}(d\theta^{\alpha})\,,~~b>0\,.~~~ \label{proprietadistrib}%
\end{equation}
The index $\alpha$ is not summed. The indices $a_{1}\dots a_{r}$ and
$\beta_{1}\dots\beta_{q}$ are anti-symmetrized, the indices $\alpha_{1}%
\dots\alpha_{s}$ are symmetrized because of the rules of the graded wedge
product:
\begin{eqnarray}
dx^{a}dx^{b}  &=&-dx^{b}dx^{a}\,,~~~dx^{a}d\theta^{\alpha}=d\theta^{\alpha
}dx^{a}\,,~~~d\theta^{\alpha}d\theta^{\beta}=d\theta^{\beta}d\theta^{\alpha
}\,,\label{wedgevari}\\
\delta(d\theta^{\alpha})\delta(d\theta^{\beta})  & =&-\delta(d\theta^{\beta
})\delta(d\theta^{\alpha})\,,\\
~~~dx^{a}\delta(d\theta^{\alpha})  &  =&-\delta(d\theta^{\alpha})dx^{a}%
\,,~~~d\theta^{\alpha}\delta(d\theta^{\beta})=\delta(d\theta^{\beta}%
)d\theta^{\alpha}\,\,.
\end{eqnarray}

As usual the module of $(p|q)$ pseudoforms is denoted by $\Omega^{(p|q)}$; if
$q=0$ or $q=m$ it is finitely generated.

It is possible to define the integral over the superspace $\mathbb{R}^{(n|m)}
$ of an \textit{integral top} form $\omega^{(n|m)}$ that can be written
locally as:
\begin{equation}
\omega^{(n|m)}=f(x,\theta)dx^{1}\dots dx^{n}\delta(d\theta^{1})\dots
\delta(d\theta^{m})\, \label{sezioneberezin}%
\end{equation}
where $f(x,\theta)$ is a superfield. By changing the $1$-forms $dx^{a}%
,d\theta^{\alpha}$ as $dx^{a}\rightarrow E^{a}=E_{m}^{a}dx^{m}+E_{\mu}%
^{a}d\theta^{\mu}$ and $d\theta^{\alpha}\rightarrow E^{\alpha}=E_{m}^{\alpha
}dx^{m}+E_{\mu}^{\alpha}d\theta^{\mu}$, we get
\begin{equation}
\omega\rightarrow\mathrm{sdet}(E)\,f(x,\theta)dx^{1}\dots dx^{n}\delta
(d\theta^{1})\dots\delta(d\theta^{m})
\end{equation}
where $\mathrm{sdet}(E)$ is the superdeterminant of the supervielbein
$(E^{a},E^{a})$.

The integral form $\omega^{(n|m)}$ can be also viewed as a superfunction
$\omega(x,\theta,dx,d\theta)$ on the \textit{odd} dual\footnote{In order to
make contact with the standard physics literature we adopt the conventions
that $d$ is an odd operator and $dx$ (an odd form) is dual to the even vector
$\frac{\partial}{\partial x}$. The same holds for the even form $d\theta$ dual
to the odd vector $\frac{\partial}{\partial\theta}.$ As clearly explained for
example in the appendix of the paper \cite{VORONOV3} if one introduces also
the natural concept of even differential (in order to make contact with the
standard definition of cotangent bundle of a manifold) our cotangent bundle
(that we consider as the bundle of one-forms) should, more appropriately, be
denoted by $\Pi T^{\ast}.$} $T^{\ast}(\mathbb{R}^{(n|m)})$ acting
superlinearly on the parity reversed tangent bundle $\Pi T(\mathbb{R}%
^{(n|m)})$, and its integral is defined as follows:
\begin{equation}
I[\omega]\equiv\int_{\mathbb{R}^{(n|m)}}\omega^{(n|m)}\equiv\int_{T^{\ast
}(\mathbb{R}^{(n|m)})=\mathbb{R}^{(n+m|m+n)}}\omega(x,\theta,dx,d\theta
)[dxd\theta~d(dx)d(d\theta)] \label{Idiomega}%
\end{equation}
where the order of the integration variables is kept fixed. The symbol
\-\ $[dxd\theta~d(dx)d(d\theta)]$ denotes the Berezin integration
\textquotedblleft measure" and it is invariant under any coordinate
transformation on $\mathbb{R}^{(n|m)}$. It is a section of the
\textit{Berezinian bundle} of $T^{\ast}(\mathbb{R}^{(n|m)})$ (a super line
bundle that generalizes the determinant bundle of a purely bosonic manifold).
The sections of the determinant bundle transform with the determinant of the
jacobian and the sections of the Berezinian with the superdeterminant of the
super-Jacobian. The berezinian bundle of $T^{\ast}\mathcal{M}^{(n|m)}$ is
always trivial but the berezinian bundle of $\mathcal{M}^{(n|m)}$ in general
is non trivial. The integrations over the fermionic variables $\theta$ and
$dx$ are Berezin integrals, and those over the bosonic variables $x$ and
$d\theta$ are Lebesgue integrals (we assume that $\omega(x,\theta,dx,d\theta)$
has compact support in the variables $x$ and it is a product of Dirac's delta
distributions in the $d\theta$ variables). A similar approach for a superform
would not be possible because the polynomial dependence on the $d\theta$ leads
to a divergent integral.

As usual, this definition can be extended to supermanifolds $\mathcal{M}%
^{(n|m)}$ by using bosonic partitions of unity.

See again Witten \cite{Witten:2012bg} for a more detailed discussion on the
symbol $[dxd\theta d(dx)d(d\theta)]$ and many other important aspects of the
integration theory of integral forms.

According to the previous discussion, if a superform $\omega^{(n|0)}$ with
form degree $n$ (equal to the bosonic dimension of the reduced bosonic
submanifold $\mathcal{M}^{(n)}\hookrightarrow\mathcal{M}^{(n|m)})$ and picture
number zero is multiplied by a $(0|m)$ integral form $\gamma^{(0|m)}$, we can
define the integral on the supermanifold of the product:%
\begin{equation}
\int_{\mathcal{M}^{(n|m)}}\omega^{(n|0)}\wedge\gamma^{(0|m)}.
\label{Iprodotto}%
\end{equation}
This type of integrals can be given a geometrical interpretation in terms of
the reduced bosonic submanifold $\mathcal{M}^{(n)}$ of the supermanifold and
the corresponding Poincar\'{e} dual (see \cite{Castellani:2015paa}).

\section{PCO's and their properties.}

In this section we recall a few definitions and useful computations about the
PCO's in our notations. For more details see \cite{Belo} and \cite{LMP}.

We start with the \textit{Picture Lowering Operators} that map cohomology
classes in picture $q$ to cohomology classes in picture $r<q.$

Given an integral form, we can obtain a superform by acting on it with
operators decreasing the picture number. Consider the following integral
operator:
\begin{equation}
\delta(\iota_{D})=\int_{-\infty}^{\infty}\exp\Big(it\iota_{D}\Big)dt
\end{equation}
where $D$ is an odd vector field with $\left\{  D,D\right\}  \neq
0\footnote{Here and in the following $\left\{  ,\right\}  $ is the
anticommutator (i.e. the graded commutator).}$ and $\iota_D$ is the
contraction along the vector $D$. The contraction $\iota_D$ is an even operator.

For example, if we decompose $D$ on a basis $D=D^{\alpha}\partial
_{\theta^{\alpha}}$, where the $D^{\alpha}$ are even coefficients and
$\left\{  \partial_{\theta^{\alpha}}\right\}  $ is a basis of the odd vector
fields, and take $\omega=\omega_{\beta}d\theta^{\beta}\in\Omega^{(1|0)}$, we
have
\begin{equation}
\iota_{D}\omega=D^{\alpha}\omega_{\alpha}=D^{\alpha}\frac{\partial\omega
}{\partial d\theta^{\alpha}}\in\Omega^{(0|0)}\,.
\end{equation}
In addition, due to $\left\{  D,D\right\}  \neq0$, we have also that
$\iota_{D}^{2}\neq0$. The differential operator $\delta(\iota_{\alpha}%
)\equiv\delta\left(  \iota_{D}\right)  $ -- with $D=\partial_{\theta^{\alpha}%
}$ -- acts on the space of integral forms as follows (we neglect the possible
introduction of derivatives of delta forms, but that generalization can be
easily done):
\begin{eqnarray}
\delta(\iota_{\alpha})\prod_{\beta=1}^{m}\delta(d\theta^{\beta})  &  =\pm
\int_{-\infty}^{\infty}\exp\Big(it\iota_{\alpha}\Big)\delta(d\theta^{\alpha
})\prod_{\beta=1\neq\alpha}^{m}\delta(d\theta^{\beta})dt\label{exaG}\\
&  =\pm\int_{-\infty}^{\infty}\delta(d\theta^{\alpha}+it)\prod_{\beta
=1\neq\alpha}^{m}\delta(d\theta^{\beta})dt=\mp i\prod_{\beta=1\neq\alpha}%
^{m}\delta(d\theta^{\beta})\nonumber
\end{eqnarray}
where the sign $\pm$ is due to the anticommutativity of the delta forms and it
depends on the index $\alpha.$ We have used also the fact that $\exp
\Big(it\iota_{\alpha}\Big)$ represents a finite translation of $d\theta
^{\alpha}$. The result contains $m-1$ delta forms, and therefore it has
picture $m-1$. It follows that $\delta(\iota_{\alpha})$ is an odd operator.

We can define also the Heaviside step operator $\Theta\left(  \iota
_{D}\right)  $ :%
\begin{equation}
\Theta\left(  \iota_{D}\right)  =\lim_{\epsilon\rightarrow0^{+}}%
-i\int_{-\infty}^{\infty}\frac{1}{t-i\epsilon}\exp\Big(it\iota_{D}\Big)dt
\label{ThetasuF}%
\end{equation}
The operators $\delta\left(  \iota_{D}\right)  $ and $\Theta\left(  \iota
_{D}\right)  $ have the usual formal distributional properties: $\iota
_{D}\delta(\iota_{D})=0$ , $\iota_{D}\delta^{\prime}(\iota_{D})=-\delta
(\iota_{D})$ and $\iota_{D}\Theta\left(  \iota_{D}\right)  =\delta(\iota_{D}).
$

In order to map cohomology classes into cohomology classes decreasing the
picture number, we introduce the operator (see \cite{Belo}):%
\begin{equation}
Z_{D}=\left[  d,\Theta\left(  \iota_{D}\right)  \right]  \label{Zeta}%
\end{equation}

In the simplest case $D=\partial_{\theta^{\alpha}}$ we have:%
\begin{equation}
Z_{\partial_{\theta^{\alpha}}}=i\delta(\iota_{\alpha})\partial_{\theta
^{\alpha}} \equiv Z_{\alpha} \label{Zetaalfa}%
\end{equation}
The operator $Z_{\alpha}$ is the composition of two operators acting on
different quantities: $\partial_{\theta^{\alpha}}$ acts only on functions, and
$\delta(\iota_{\alpha})$ acts only on delta forms.

In order to further reduce the picture we simply iterate operators of type
$Z$. An alternative description of $Z$ in terms of the Voronov integral
transform can be found in \cite{LMP}.

The $Z$ operator is in general not invertible but it is possible to find a non
unique operator $Y$ such that $Z\circ Y$ is an isomorphism in the cohomology.
These operators are the called \textit{Picture Raising Operators. }The
operators of type $Y$ are non trivial elements of the de Rham cohomology.

We apply a PCO of type $Y$ on a given form by taking the graded wedge product;
given $\omega$ in $\Omega^{(p|q)}$, we have:
\begin{equation}
\omega\overset{Y}{\longrightarrow}\omega\wedge Y\in\Omega^{\left(
{p|q+1}\right)  }\,, \label{PCOc}%
\end{equation}
Notice that if $q=m$, then $\omega\wedge Y=0$. In addition, if $d\omega=0$
then $d(\omega\wedge Y)=0$ (by applying the Leibniz rule), and if $\omega\neq
dK$ then it follows that also $\omega\wedge Y\neq dU$ where $U$ is a form in
$\Omega^{(p-1|q+1)}$. So, given an element of the cohomogy $H^{(p|q)}$, the
new form $\omega\wedge Y$ is an element of $H^{(p|q+1)}.$

For a simple example in $\mathbb{R}^{(1|1)}$ we can consider the PCO
$Y=\theta\delta\left(  d\theta\right)  $, corresponding to the vector
$\partial_{\theta}$; we have $Z\circ Y=Y\circ Z=1$

More general forms for $Z$ and $Y$ can be constructed, for example starting
with the vector $Q=\partial_{\theta}+\theta\partial_{x}.$

For example, if $\varphi=g(x)\theta dx\delta(d\theta)$ is a generic top
integral form in $\Omega^{(1|1)}\left(  \mathbb{R}^{(1|1)}\right)  ,$ the
explicit computation using the formula $Z=[d,\Theta(\iota_{Q})]$ is:
\begin{eqnarray}
Z_{Q}[\varphi]  &  =d[\Theta(\iota_{Q})\varphi]=d\Big[\Theta(\iota
_{Q})g(x)\theta dx\delta(d\theta)\Big]\\
&  =d\Big[\lim_{\epsilon\rightarrow0^{+}}-i\int_{-\infty}^{\infty}\frac
{1}{t-i\epsilon}g(x)\theta dx\delta(d\theta
+it)dt\Big]=\nonumber\label{ThetasuFi}\\
&  =d\left[  -\frac{g(x)\theta dx}{d\theta}\right]  =-g(x)dx\,.\nonumber
\end{eqnarray}
The last expression is clearly closed. Note that in the above computations we
have introduced formally the inverse of the (commuting) superform $d\theta.$
Using a terminology borrowed from superstring theory we can say that, even
though in a computation we need an object that lives in the \textit{Large
Hilbert Space}, the result is still in the \textit{Small Hilbert Space}.

Note that the negative powers of the superform $d\theta$ are well defined only
in the complexes of superforms (i.e. in picture $0)$. In this case the inverse
of the $d\theta$ and its powers are closed and exact and behave with respect
to the graded wedge product as \textit{negative degree} superforms of picture
$0$. In picture $\neq0$ negative powers are not defined because of the
distributional relation $d\theta\delta\left(  d\theta\right)  =0.$

A PCO of type $Y$ invariant under the rigid supersymmetry transformations
(generated by the vector $Q$) $\delta_{\epsilon}x=\epsilon\theta$ and
$\delta_{\epsilon}\theta=\epsilon$ is, for example, given by:%
\begin{equation}
Y_{Q}\mathbb{=}(dx+\theta d\theta)\delta^{\prime}(d\theta)
\end{equation}
We have:
\begin{equation}
Y_{Q}Z_{Q}[\varphi]=-g(x)dx\wedge(dx+\theta d\theta){}\delta^{\prime}%
(d\theta)=g(x)\theta dx\delta(d\theta)=\varphi\,.
\end{equation}

\section{Rheonomic Sigma Model}

We consider a flat complex superspace with bosonic coordinates $(z=z^{++}, \bar z = z^{--})$
and Grassmanian coordinates $(\theta =\theta^+, \bar \theta = \theta^{-})$. The charges $\pm$ are assigned according to the
transformation properties of the coordinates $z, \theta$ under the Lorentz group $SO(1,1)$. The latter being unidimensional, the irreducible representations are parametrized by their charges
\begin{eqnarray}
\label{lorA}
x^{\pm\pm} \rightarrow e^{\pm i \theta} x^{\pm\pm} \,, ~~~~~
\theta^\pm \rightarrow e^{\pm\frac{ i \theta}{2} } \theta^{\pm}\,. 
\end{eqnarray}

We introduce the 
differentials $(dx^{\pm\pm}, d\theta^\pm)$ and the flat supervielbeins
\begin{equation}
V^{\pm\pm}=dz^{\pm\pm}+\theta^\pm d\theta^\pm\,,~~~~~~
\psi^\pm=d\theta^\pm\,,
\end{equation}
invariant under the rigid supersymmetry $\delta\theta^\pm=\epsilon^\pm$ and 
$\delta x^{\pm\pm}=\epsilon^\pm\theta^\pm$. They
satisfy the MC algebra
\begin{equation}
dV^{\pm\pm}=\psi^\pm\wedge\psi^\pm\,,~~~~~~
d\psi^\pm=0\,.
\end{equation}
We first consider the non-chiral multiplet. This is described by a superfield
$\Phi$ with the decomposition
\begin{eqnarray}
\Phi &=&\phi+\lambda\theta^++\bar{\lambda}{\theta}^-+f\theta^+ {\theta}^-
\nonumber\label{oneB}\\
W  &  =&D_+ \Phi\,,\nonumber\\
\bar{W}  &  =&{D}_- \Phi\,,\nonumber\\
F  &  =&D_- D_+\Phi\,.
\end{eqnarray}
where $D_+=\partial_{\theta^{+}}-\frac{1}{2}\theta^+\partial_{++}$ and 
${D}_-=\partial_{{\theta}^-}-\frac{1}{2}{\theta}^-{\partial_{--}}$ (with
$\partial_{++}=\partial_{z^{++}}$ and 
$\partial_{--}=\partial_{{z}^{--}}$). They satisfy
the algebra $D^{2}_+=-\partial_{++}$ and 
${D}^{2}_-=- {\partial}_{--}$ and
anticommute $D_- {D}_+ + {D}_+ D_-=0$. The component fields 
$\phi,\lambda,\bar{\lambda}$ and $f$ are spacetime fields and they depend only upon
$z^{\pm\pm}$. On the other hand, $(\Phi,W,\bar{W},F)$ are the superfields
whose first components are the components fields. $W$ and $\bar{W}$ are
anticommuting superfields.

Computing the differential of each superfield we have the following
relations:
\begin{eqnarray}
d\Phi &=&V^{++}\partial_{++}\Phi+
{V}^{--}{\partial}_{--}\Phi+\psi^+ W+{\psi}^-\bar
{W}\,,\nonumber\label{oneC}\\
dW  &=&V^{++}\partial_{++} W+ {V}^{--} {\partial}_{--} W-
\psi^+\partial_{++}\Phi+{\psi}^- F\,,\nonumber\\
d\bar{W}  &=&V^{++} \partial_{++}\bar{W}+
{V}^{--} {\partial}_{--}\bar{W}- {\psi}^-%
{\partial}_{--}\Phi-\psi^+ F\,,\nonumber\\
dF  &=&V^{++}\partial_{++} F + {V}^{--}
\bar{\partial}_{--}F+\psi^+\partial_{++}\bar{W}-
{\psi}^- {\partial}_{--} W\,,
\end{eqnarray}
The last field $F$ is the auxiliary field and therefore it vanishes when the
theory is on-shell. Before writing the rheonomic Lagrangian for the multiplet,
we first write the equations of motion. If we set $F=0$, then we see from the
last equation that
\begin{equation}
\partial_{++}\bar{W}=0\,,~~~~~{\partial}_{--}W=0\,.
\end{equation}
They implies that the superfield $W$ is holomorphic $W=W(z)$ and the
superfield $\bar{W}$ is anti-holomorphic. Then, we can write eqs. (\ref{oneC})
with these constraints:
\begin{eqnarray}
\label{cicA}
d\Phi &  =&V^{++}\partial_{++}\Phi+{V}^{--} {\partial}_{--}\Phi+\psi^+ W+
{\psi}^-\bar{W}\,,\nonumber\label{oneE}\\
dW  &  =&V^{++}\partial_{++} W-\psi^+ \partial_{++}\Phi\,,\nonumber\\
d\bar{W}  &  =&{V}^{--} {\partial}_{--}\bar{W}- {\psi}^- {\partial}_{--}\Phi\,,
\end{eqnarray}
The consistency of the last two equations $(d^{2}=0)$, implies that
$\partial_{++} {\partial}_{--}\Phi=0$, Then, we get that the rheonomic equations
(\ref{oneC}) are compatible with the set of the equations of motion
\begin{equation}
\partial_{++} 
{\partial}_{--} \Phi=0\,,~~~~~~~
\partial_{++}\bar{W}=0\,,~~~~~~~
{\partial}_{--}W=0\,,~~~~~~~F=0\,. \label{oneE}%
\end{equation}
which are the free equations of $D=2$ multiplet. The Klein-Gordon equation in
$D=2$ implies that the solution $\Phi=\Phi_{h}(z)+\Phi_{\bar{h}}(\bar{z})$ is
splitted into holomorphic and anti-holomorphic parts and therefore we get the
on-shell matching of the degrees of freedom. In particular we can write
on-shell holomorphic and anti-holomorphic superfields
\begin{equation}
\Phi_{h}(z)=\phi(z)+\lambda(z)\theta^+\,,~~~~~
\Phi_{\bar{h}}(\bar{z})=\phi
(\bar{z})+\bar{\lambda}(\bar{z}) {\theta}^-\,,~~~~~
\end{equation}
factorizing into left- and right-movers.

Let us now write the action. We introduce two additional superfields $\xi$ and
$\bar{\xi}$. Then, we have \cite{cube}
\begin{eqnarray}
\mathcal{L}^{(2|0)}  &&  =(\xi V^{++}+\bar{\xi} V^{--})\wedge(d\Phi-\psi^+ W-
{\psi}^-\bar{W})+\left(  \xi\bar{\xi}+\frac{F^{2}}{2}\right)  V^{++}\wedge{V}^{--}
\nonumber\label{oneH}\\
&&  +WdW\wedge V^{++}-\bar{W}d\bar{W}\wedge{V}^{--}-d\Phi\wedge(W\psi^+-\bar{W}%
{\psi}^-)-W\bar{W}\,\psi^+\wedge{\psi}^- \nonumber \\%
\end{eqnarray}
The equations of motion are given by
\begin{eqnarray}
&&  V^{++}\wedge(d\Phi-\psi^+ W- {\psi}^- \bar{W})+\bar{\xi}V^{++}\wedge {V}^{--}%
=0\,,\nonumber\label{oneHA}\\
&&  {V}^{--}\wedge(d\Phi-\psi^+ W- {\psi}^- \bar{W})+\xi V^{++}\wedge {V}^{--}=0\,,\nonumber\\
&&  (\xi V^{++}+\bar{\xi}{V}^{--})\psi^++2dW\wedge V^{++}-W\psi^+\wedge\psi^++d\Phi\wedge
\psi^+-\bar{W}\psi^+\wedge{\psi}^-=0\,,\nonumber\\
&&  (\xi V^{++}+\bar{\xi} {V}^{--}){\psi}^--2d\bar{W}\wedge V^{--}+\bar{W}{\psi}^-
\wedge{\psi}^{-}-d\Phi\wedge{\psi}^{-}+W\psi^+\wedge{\psi}^-=0\,,\nonumber\\
&&  d(\xi V^{++}+\bar{\xi}{V}^{--})+dW\psi^+-d\bar{W}{\psi}^-=0\,,\nonumber\\
&&  F=0\,.
\end{eqnarray}
They imply the on-shell differentials (\ref{cicA}), the equations of motion
(\ref{oneE}), and the relations
\begin{equation}
\xi=\partial_{++}\Phi\,,~~~~~\bar{\xi}=-{\partial}_{--}\Phi\,. \label{abbotA}%
\end{equation}
expressing the additional auxiliary fields $\xi$ and $\bar{\xi}$ in terms of
$\Phi$. It is easy to check that they are consistent: acting with $d$ on the
third and on the fourth equations, and using the fifth equation, one gets a
trivial consistency check. In the same way for all the others.

The action is a $(2|0)$ superform, it can be verified that it is closed by
using only the algebraic equations of motion for $\xi$ and $\bar{\xi}$, which
are solved in (\ref{abbotA}) and using the curvature parametrization
$d\Phi,dW,d\bar{W}$ and $dF$ given in (\ref{oneC}). Note that those equations
are off-shell parametrizations of the curvatures and therefore they do not
need the equations of motion of the lagrangian (\ref{oneH}).

\subsection{Sigma Model on Supermanifolds}

To check whether this action leads to the correct component action we use the
PCO ${\mathbb{Y}}^{(0|2)}=\theta^+ {\theta}^-\delta(\psi^+)\delta({\psi}^-)$.
Then we have\footnote{We denote by $\int_{\mathcal{M}}$ the integral of an
integral form on the supermanifold, by $\int[d^{2}zd^{2}\theta]$ the Berezin
integral on the superspace and by $\int d^{2}z$ the usual integral on the
reduced bosonic submanifold.}
\begin{eqnarray}
S  &  =&\int_{\mathcal{S}\Sigma}\mathcal{L}^{(2|0)}\wedge{\mathbb{Y}}^{(0|2)}%
\label{oneI}\\
&  =&\int d^{2}z\left[  (\xi_{0}dz^{++}+\bar{\xi}_{0}d {z}^{--})\wedge d\phi+\left(
\xi_{0}\bar{\xi}_{0}+\frac{f^{2}}{2}dz^{++}\wedge d{z}^{--}\right)  \right. \nonumber \\
&&~~~~~~~~ +\left. \lambda
d\lambda\wedge dz^{++}+\bar{\lambda}d\bar{\lambda}\wedge d{z}^{--}\right]
\nonumber
\end{eqnarray}
where $\xi_{0}$ and $\bar{\xi}_{0}$ are the first components of the
superfields $\xi$ and $\bar{\xi}$. Eliminating $\xi_{0}$ and $\bar{\xi}_{0}$
one finds the usual equations of motion for the $D=2$ free sigma model.

Choosing a different PCO of the form\footnote{This form of the PCO recalls the
string theory PCO $c\delta^{\prime}(\gamma)$ where $c$ is the diffeormophism
ghost and $\gamma$ is the superghost.}
\begin{equation}
{\mathbb{Y}}^{(0|2)}=V^{++} \delta'(\psi^+)\wedge {V}^{--} \delta'
({\psi}^{-})\,,
\end{equation}
which has again the correct picture number and is cohomologous to the previous
one, leads to the superspace action (listing only the relevant terms)
\begin{eqnarray}
S  &  =&\int_{\mathcal{M}}\left[  W\bar{W}\psi^+\wedge{\psi}^--d\Phi
\wedge(W\psi^++\bar{W}{\psi}^-)\right]  \wedge V^{++}\delta'(\psi^+)\wedge
{V}^{--}\delta'({\psi}^-)\nonumber\label{oneM}\\
&  =&\int_{\mathcal{M}}\Big(W\bar{W}-[({\iota}_-d\phi)W+(\iota_+ d\phi)\bar
{W}]\Big)V^{++}\wedge {V}^{--}\wedge\delta(\psi^+)\delta({\psi}^-)\,.
\end{eqnarray}
where $\iota_{\pm}$ are the derivatives with respect to $\psi^\pm$. 
The contractions give $\iota_+ d\Phi=D_+\Phi$ and ${\iota}_-d\Phi=\bar{D}_-\Phi$.
Then we get the superspace action
\begin{equation}
S=\int[d^{2}zd^{2}\theta]\left(  W\bar{W}- {D}_-\Phi W-D_+\Phi\bar{W}\right)\,. 
\end{equation}
The equation of motion are $W=D_+\Phi$ and $\bar{W}={D}_-\Phi.$ Hence we
obtain the usual $D=2$ superspace free action in a flat background:%
\begin{equation}
S=\int[d^{2}zd^{2}\theta]D_+\Phi {D}_-\Phi\,. 
\end{equation}

\section{Geometry of $OSp(1|2)/SO(1,1)$}

Let us consider the coset $OSp(1|2)/SO(1,1)$. The MC equations can be easily
computed by using the notation $V^{\pm\pm},\psi^{\pm}$ for the MC forms and
$\nabla$ for the $SO(1,1)$ covariant derivate. The MC forms $V^{\pm\pm}$ have
charge $\pm2$, while $\psi^{\pm}$ have charge $\pm1$. Then, we have
\[
\nabla V^{++}=\psi^{+}\wedge\psi^{+}\,,~~~~~\nabla V^{--}=\psi^{-}\wedge
\psi^{-}\,,~~~~~\nabla\psi^{+}=V^{++}\wedge\psi^{-}\,,~~~~~\nabla\psi
^{-}=-V^{--}\wedge\psi^{+}\,,~~~~~
\]
Computing the Bianchi identities, we have
\begin{eqnarray}
&  \nabla^{2}V^{\pm\pm}=\pm V^{\pm\pm}\wedge R^{(2|0)}\,,~~~~~~\nabla^{2}%
\psi^{\pm}=\pm\psi^{\pm}\wedge R^{(2|0)}\,,~~~~~~\nonumber\label{OsB}\\
&  R^{(2|0)}=-V^{++}\wedge V^{--}+\,\psi^{+}\wedge\psi^{-}\,,~~~~~~\nabla
R^{(2|0)}=0\,,
\end{eqnarray}
All the expressions have been constructed to respect the charge assignements.
The superform $R^{(2|0)}$ is neutral and invariant. 


The volume form is computed by observing that
\begin{equation}
\mathrm{Vol}^{(2|2)}=V^{++}\wedge V^{--}\delta(\psi^{+})\delta(\psi
^{-})=\mathrm{Sdet}(E)d^{2}z\delta^{2}(d\theta)\,.
\end{equation}
where $\mathrm{Sdet}(E)$ is the Berezinian of the supervielbein $E$ of the
super-coset manifold ${OSp}(1|2)/SO(1,1)$. It is susy invariant and it is
closed. This can be checked by observing that
\begin{eqnarray}
d\mathrm{Vol}^{(2|2)}=\nabla\mathrm{Vol}^{(2|2)}  &&  =\left(  \nabla
V^{++}\wedge V^{--}-V^{++}\nabla V^{--}\right)  \delta(\psi^{+})\delta
(\psi^{-})\label{OsD}\\
&&  +\, V^{++}\wedge V^{--}\left(  \nabla\delta(\psi^{+})\delta(\psi^{-}%
)-\delta(\psi^{+})\nabla\delta(\psi^{-})\right) \nonumber\\
&&  +\left(  \psi^{+}\wedge\psi^{+}\ \wedge V^{--}-V^{++}\wedge\psi^{-}%
\wedge\psi^{-}\right)  \delta(\psi^{+})\delta(\psi^{-})\nonumber\\
&&  +\, V^{++}\wedge V^{--}\Big(V^{++}\wedge\psi^{-}\delta^{\prime}(\psi
^{+})\delta(\psi^{-})+\delta(\psi^{+})V^{--}\wedge\psi^{+}\delta^{\prime}%
(\psi^{-})\Big)=0\nonumber
\end{eqnarray}
The first equality follows from the neutrality of the volume integral form
$\mathrm{Vol}^{(2|2}$ and therefore, we can use the covariant derivative
instead of the differential $d$. The covariant differential $\nabla$ acts as
derivative, and this leads to the last two lines. The third line cancels
because of the Dirac delta functions multiplied by $\psi^{\pm}$, the fourth
line vanishes since $V^{\pm\pm}\wedge V^{\pm\pm}=0$.

The relevant set of pseudo-forms are contained in the rectangular diagram in
fig. (\ref{twocomplex}). The vertical arrows denote a PCO which increases the picture number. 
There are additional sets outsides the present
rectangular set, but are unessential for the present discussion since
they do not contain non trivial cohomology classes (see
\cite{Catenacci:2010cs}).

\begin{figure}[t]
\begin{center}%
\begin{tabular}
[c]%
{@{\hskip1.5mm}c@{\hskip1mm}c@{\hskip1.5mm}c@{\hskip1.5mm}c@{\hskip1.5mm}c@{\hskip1.5mm}}
&  &  &  & \\[-0.34cm]%
$\Omega^{(0|0)}$ & $\overset{{\nabla}}{\longrightarrow}$ & $\Omega^{(1|0)}$ &
$\overset{{\nabla}}{\longrightarrow}$ & $\Omega^{(2|0)}$\\[0.06cm]%
{\large {$\downarrow$}} &  & {\large {$\downarrow$}} &  & {\large {$\downarrow
$}}\\[-0.05cm]%
$\Omega^{(0|1)}$ & $\overset{{\nabla}}{\longrightarrow}$ & $\Omega^{(1|1)}$ &
$\overset{{\nabla}}{\longrightarrow}$ & $\Omega^{(2|1)}$\\[-0.15cm]%
{\large {$\downarrow$}} &  & {\large {$\downarrow$}} &  & {\large {$\downarrow
$}}\\[0.05cm]%
$\Omega^{(0|2)}$ & $\overset{{\nabla}}{\longrightarrow}$ & $\Omega^{(1|2)}$ &
$\overset{{\nabla}}{\longrightarrow}$ & $\Omega^{(2|2)}$\\[0.08cm]
&  &  &  &
\end{tabular}
\end{center}
\par
\vskip -0.2cm\caption{{\small The pseudoform complexes.}}%
\label{twocomplex}%
\end{figure}

Let us discuss the relevant cohomology spaces.
\begin{eqnarray}
&  H^{(0|0)}=\{1\}\,,\nonumber\label{OsE}\\
&  H^{(0|1)}=\{V^{++}\wedge\delta^{\prime}(\psi^{+}),V^{--}\delta^{\prime
}(\psi^{-})\}\,,\nonumber\\
&  H^{(0|2)}=\{V^{++}\wedge V^{--}\delta^{\prime}(\psi^{+})\delta^{\prime
}(\psi^{-})\}\,,\nonumber\\
&  H^{(2|0)}=\{V^{++}\wedge V^{--} - \psi^{+}\wedge\psi^{-}\}\,,\nonumber\\
&  H^{(2|1)}=\{V^{++}\wedge\psi^{-}\delta(\psi^{+}),V^{--}\wedge\psi^{+}%
\delta(\psi^{-})\}\,,\nonumber\\
&  H^{(2|2)}=\{V^{++}\wedge V^{--}\delta(\psi^{+})\delta(\psi^{-})\}\,,
\end{eqnarray}

It is easy to check the closure of all generators. In addition, all generators
are neutral, for instance in $V^{++}\wedge\delta^{\prime}(\psi^{+})$ the
charge $+2$ is compensated by the negative charge $-1$ of $\delta(\psi^{+})$
and the negative charge of the derivative of the delta form. It can be shown
that
\begin{eqnarray}
&  H^{(2|0)}\wedge H^{(0|2)}=(V^{++}\wedge V^{--}- \psi^{+}\wedge\psi
^{-})\wedge V^{++}\wedge V^{--}\delta^{\prime}(\psi^{+})\delta^{\prime}%
(\psi^{-})\nonumber\label{OsF}\\
&  \longrightarrow H^{(2|2)}=V^{++}\wedge V^{--}\delta(\psi^{+})\delta
(\psi^{-})
\end{eqnarray}
This equation is rather suggestive. If we consider the cohomology class in
$H^{(0|2)}$ as the total PCO $\mathbb{Y}^{(0|2)}$ and if we consider the
cohomology class in $H^{(2|0)}$ as the K\"{a}hler form $K^{(2|0)}$ of our
complex supermanifold, we find
\begin{equation}
\mathbb{Y}^{(0|2)}\wedge R^{(2|0)}=\mathbb{Y}^{(0|2)}\wedge K^{(2|0)}%
=\mathrm{Vol}^{(2|2)}\,,
\end{equation}
which is the super-Liouville form.

The PCO operator $\mathbb{Y}^{(0|2)}$ easily factorizes as $\mathbb{Y}^{(0|2)}
= \mathbb{Y}^{(0|1)}_{+} \wedge\mathbb{Y}^{(0|1)}_{-}$ with $\mathbb{Y}%
^{(0|1)}_{\pm}= V^{\pm\pm} \wedge\delta^{\prime}(\psi^{\pm})$. This
factorization is very useful for the equations below.


Finally, we want to show that acting with the PCO $Z$ we can map the volume
form $\mathrm{Vol}^{(2|2)}$ into the K\"{a}hler form $K^{(2|0)}$. For that we
define the PCO's
\begin{equation}
Z_{+}^{(0|-1)}=[d,\Theta(\iota_{+})]\,,~~~~~~Z_{-}^{(0|-1)}=[d,\Theta
(\iota_{-})]\,.~~~~~~
\end{equation}
Acting with the first one on $\mathrm{Vol}^{(2|2)}$ (and using the fact that
$d\mathrm{Vol}^{(2|2)}=0$, we have
\begin{eqnarray}
Z_{+}^{(0|-1)}\Big(V^{++}\wedge V^{--}\delta(\psi^{+})\delta(\psi^{-})\Big)
&  =d\Big(\Theta(\iota_{+})V^{++}\wedge V^{--}\delta(\psi^{+})\delta(\psi
^{-})\Big)\nonumber\label{OsN}\\
&  =d\Big(V^{++}\wedge V^{--}\frac{1}{\psi^{+}}\delta(\psi^{-}%
)\Big)\nonumber\\
&  =\psi^{+}\wedge V^{--}\delta(\psi^{-})\in H^{(2|1)}%
\end{eqnarray}
It can be noticed that the final expression is chargeless, it is $d$-closed
and it is expressed in terms of supersymmetric invariant quantities.
Furthermore, in the first step of the computation we have used objects in the
Large Hilbert Space, but the final result is again in the Small Hilbert Space
(there are no inverse of $\psi$'s)\footnote{Note the important point:
expressions like $\frac{1}{\psi^{+}}\delta(\psi^{-})$ or $\frac{1}{\psi^{-}%
}\delta(\psi^{+})$ are well defined.}.

Let us act with the second PCO, $Z_{-}^{(0|-1)}=[d,\Theta(\iota_{-})]$. Again,
we use the fact that the result of (\ref{OsN}) is $d$-closed. We have
\begin{eqnarray}
Z_{-}^{(0|-1)}\Big(\psi^{+}\wedge V^{--}\delta(\psi^{-})\Big)  &
=&d\Big(\Theta(\iota_{-})\psi^{+}\wedge V^{--}\delta(\psi^{-}%
)\Big) \label{OsP}\\
&  =&d\Big(\psi^{+}\wedge V^{--}\frac{1}{\psi^{-}}\Big)=V^{++}\wedge
V^{--}-\psi^{+}\wedge\psi^{-}\in H^{(2|0)} \nonumber
\end{eqnarray}
Again, in the intermediate steps we have expressions living into the Large
Hilbert Space, but the final result is in the Small Hilbert Space and it is
polynomial in the MC forms. This clearly shows how to act with the PCO's on
the cohomology classes mapping from cohomology to cohomology. The same result
is obtained by exchanging the two PCO's. The final result is
\begin{equation}
Z_{-}^{(0|-1)}Z_{+}^{(0|-1)}\mathrm{Vol}^{(2|2)}=\frac{1}{4}R^{(2|0)}
\label{OSPP}%
\end{equation}
mapping the volume form into the curvature of the manifold.

\section{D=2 Supergravity}

As is well know, there are no dynamical graviton and gravitino in $2$
dimensions, nonetheless the geometric formulation of supergravity is
interesting and it is relevant in the present work. The definitions are
\begin{eqnarray}
&&  T^{\pm\pm}=\nabla V^{\pm\pm}\pm\frac{1}{2}\psi^{\pm}\wedge\psi^{\pm
}\,,~~~~~~\label{defA}\\
&&  \rho^{\pm}=\nabla\psi^{\pm}\,,~~~~~~\nonumber\\
&&  R=d\omega\,,\nonumber
\end{eqnarray}
where $\omega$ is the $SO(1,1)$ spin connection. These curvatures satisfy the
following Bianchi identities
\begin{eqnarray}
&&  \nabla T^{\pm\pm}=\mp2R\wedge V^{\pm\pm}\mp\rho^{\pm}\wedge\psi^{\pm
}~~~~~~\label{defB}\\
&&  \nabla\rho^{\pm}=\mp R\wedge\psi^{\pm}\,,~~~~~~\nonumber\\
&&  \nabla R=0\,.\nonumber
\end{eqnarray}

The Bianchi identities can be solved by the following parametrization
\begin{eqnarray}
T^{\pm\pm} &= &  \,0\,,~~~~~\label{defC}\\
\rho^{\pm} &= &  \,4D_{\mp}E\,V^{++}\wedge V^{--}-2E\,\psi^{\mp}\wedge
V^{\pm\pm}\,,~~~~~\nonumber\\
R&=  &  \,-4(D_{+}D_{-}E+E^{2})\,V^{++}\wedge V^{--}~~~~~\nonumber\\
&&  \,-2D_{+}E\,V^{++}\wedge\psi^{-}+2D_{-}E\,V^{--}\wedge\psi^{+}+E\,\psi
^{+}\wedge\psi^{-}\,,\nonumber
\end{eqnarray}
where $E$ is a generic superfield $E(x,\theta)=E_{0}(x)+E_{+}(x)\theta
^{+}+E_{-}(x)\theta^{-}+E_{1}(x)\theta^{+}\theta^{-}$. There are no dynamical
constraint on $E(x,\theta)$ since there are no equations of motion.
Nonetheless if we impose that
\begin{equation}
D_{+}E=D_{-}E=0 \label{defD}%
\end{equation}
we immediately get $\partial_{++}E=\partial_{--}E=0$, and therefore $E=const$.
If we set $E=\Lambda$ we get the well-known anti-de-Sitter solution
\begin{eqnarray}
T^{\pm\pm}&=  &  \,0\,,~~~~~\label{defE}\\
\rho^{\pm}&=  &  -2\Lambda\,\psi^{\mp}\wedge V^{\pm\pm}\,,~~~~~\nonumber\\
R&=  &  \,-4\Lambda^{2}\,V^{++}\wedge V^{--}+\Lambda\,\psi^{+}\wedge\psi
^{-}\,,\nonumber
\end{eqnarray}
describing the coset space $\mathrm{Osp}(1|2)/\mathrm{SO}(1,1)$.

Going back to a generic $E$, we consider the volume form
\begin{equation}
\mathrm{Vol}^{(2|2)}=E\,V^{++}\wedge V^{--}\delta(\psi^{+})\delta(\psi^{-})
\label{defF}%
\end{equation}
which is closed since it is a top integral form 
Now we
act with the PCO $Z_{+}=[d,\Theta(\iota_{+})]$ and we get
\begin{eqnarray}
Z_{+}\mathrm{Vol}^{(2|2)}  &  =d\Big(\Theta(\iota_{+})\mathrm{Vol}%
^{(2|2)}\Big)=\Big(D_{+}E\,V^{++}\wedge V^{--}-\frac{1}{2}E\,V^{--}\wedge
\psi^{+}\Big)\delta(\psi^{-})\nonumber\label{defG}\\
&  =\frac{1}{4}\rho^{-}\delta(\psi^{-})\,.
\end{eqnarray}
Notice that the Dirac delta's do not carry any charges and the PCO $Z_{+}$ has
negative charge. Therefore, the result is consistent. In addition, the r.h.s.
is closed, as can be easily verified by using the Bianchi identites:
\begin{eqnarray}
\nabla\Big(\rho^{-}\delta(\psi^{-})\Big)  &  =(\nabla\rho^{-})\delta(\psi
^{-})-\rho^{-}\delta^{\prime}(\psi^{-})\wedge\nabla\psi^{-}%
\nonumber\label{defH}\\
&  =R\wedge\psi^{-}\delta(\psi^{-})-\rho^{-}\delta^{\prime}(\psi^{-}%
)\wedge\rho^{-}=0
\end{eqnarray}
since $\rho^{-}\wedge\rho^{-}=0$ and $\psi^{-}\delta(\psi^{-})=0$.

Let us now act with the second PCO $Z_{-}$. There are two ways to perform the
computation: either using the complete expression given in the first line of
(\ref{defG}) or using the Bianchi identities. With the second proposal we
observe:
\begin{eqnarray}
Z_{-}Z_{+}\mathrm{Vol}^{(2|2)}  &  =Z_{-}\Big(\frac{1}{4}\rho^{-}\delta
(\psi^{-})\Big)=\frac{1}{4}\nabla\Big(\Theta(\iota_{-})\rho^{-}\delta(\psi
^{-})\Big)\nonumber\label{defI}\\
&  =\frac{1}{4}\nabla\Big(\frac{\rho^{-}}{\psi^{-}}\Big)=\frac{1}{4}R
\end{eqnarray}
Notice that acting both with $Z_{-}$ and $Z_{+}$ the total charge is zero as
for $R$. The result is closed, $dR=0$ , and it confirms the formula obtained
for the curved rigid supermanifold (\ref{OSPP}). The result (\ref{defI}) is
valid for any superfield $E$.

The PCO $Y$ are defined as in the flat case
\begin{equation}
Y^{+}=V^{++}\delta^{\prime}(\psi^{+})\,,~~~~~Y^{-}=V^{--}\delta^{\prime}%
(\psi^{-})\,,~~~~~ \label{YYA}%
\end{equation}
We can easily check their closure:
\begin{eqnarray}
\nabla Y^{+}  &  =&\nabla V^{++}\delta^{\prime}(\psi^{+})+V^{++}\delta
^{\prime\prime}(\psi^{+})\nabla\psi^{+}\label{YYB}\\
&  =&\left(  T^{++}+\frac{1}{2}\psi^{+}\wedge\psi^{+}\right)  \delta^{\prime
}(\psi^{+})+V^{++}\delta^{\prime\prime}(\psi^{+})\Big(4D_{\mp}E\,V^{++}\wedge
V^{--}-2E\,\psi^{-}\wedge V^{++}\Big)\nonumber\\
&  =&T^{++}\delta^{\prime}(\psi^{+})\nonumber
\end{eqnarray}
Therefore, it is closed if $T^{++}=0$. In the same way we get for $Y^{-}$.
Finally, we observe that
\begin{equation}
Y^{+}\wedge Y^{-}\wedge R=\mathrm{Vol}^{(2|2)} \label{YYC}%
\end{equation}
This can also be obtained by observing that
\begin{eqnarray}
Z_{+}Y^{+}  &  =&d\Big(\Theta(\iota_{+})V^{++}\delta^{\prime}(\psi
^{+})\Big)=d\left(  \frac{V^{++}}{\psi^{+}\wedge\psi^{+}}\right)
\nonumber\label{YYD}\\
&  =&\frac{1}{2}+2\frac{V^{++}\wedge\rho^{+}}{\psi^{+}\wedge\psi^{+}\wedge
\psi^{+}}=\frac{1}{2}%
\end{eqnarray}
since $V^{++}\wedge\rho^{+}=0$. In the same way, $Z_{-}Y^{-}=1/2$. These
equations are valid for any $E$.
Using eq. (\ref{YYC}), one can define an integral over the supermanifold: 
\begin{eqnarray}
\label{YYE}
\int_{{\cal S}\Sigma} R \wedge Y^+ \wedge Y^- = 
\int_{{\cal S}\Sigma} {\rm Vol}^{(2|2)} = \int_{\Sigma} D_{+}D_{-} E \, V^{++}\wedge V^{--} 
\end{eqnarray}
that might be interpreted as the Euler characteristic for supermanifolds.  
\section{Conclusions}

To complete the program, one has to use the PCO's (\ref{YYA}), for a generic background $E$ 
to rewrite the action (\ref{actF}) in that background. Choosing a different PCO gives an equivalent 
string sigma model with different manifest supersymmetry. 

Finally, we would like to point out the relation between the PCO used in the action, and 
the conventional PCO used for correlation computations in string theory. The latter can be 
written as follows
\begin{eqnarray}
\label{strA}
Y = c^{++}\delta'(\gamma^+)\,, ~~~~~
\bar Y = c^{--} \delta'(\gamma^-) 
\end{eqnarray}
 for the left- and right-moving sector, where $c^{\pm\pm}$ are the Einstein's ghosts and 
 $\gamma^\pm$ are the superghosts. They should be compared with 
 (\ref{YYA}). We further notice that BRST transformations of the D=2 supervielbeins 
 $V^{\pm\pm}, \psi^\pm$  are 
 given by 
 \begin{eqnarray}
\label{strB}
Q V^{\pm\pm} = d c^{\pm\pm} + \dots\,, ~~~~
Q \psi^\pm = d \gamma^\pm + \dots
\end{eqnarray}
where the ellipsis denotes non-linear terms, and therefore there should be a relation between the two 
types of PCO. We leave this to further investigations. 
  
\section*{Acknowledgments}
{The paper is supported in part by Fondi Ricerca Locale (ex 60\%) Theoretical Physics. 
The authors would like to acknowledgement Leonardo Castellani and Pietro Fr\'e for useful dicussions 
and comments.} 

\vfill
\eject


\end{document}